\documentclass[12pt]{iopart}
\usepackage{graphicx} 
\usepackage[footnotesize,figbotcap]{subfigure} 
\usepackage{cite} 

\begin{document}

\title{High $p_T$ Triggered $\Delta\eta,\Delta\phi$ Correlations over a Broad Range in $\Delta\eta$    }
\author{\large Edward Wenger for the PHOBOS collaboration}
%
%
\author{
B.Alver$^4$,
B.B.Back$^1$,
M.D.Baker$^2$,
M.Ballintijn$^4$,
D.S.Barton$^2$,
R.R.Betts$^6$,
A.A.Bickley$^7$,
R.Bindel$^7$,
W.Busza$^4$,
A.Carroll$^2$,
Z.Chai$^2$,
V.Chetluru$^6$,
M.P.Decowski$^4$,
E.Garc\'{\i}a$^6$,
T.Gburek$^3$,
N.George$^2$,
K.Gulbrandsen$^4$,
C.Halliwell$^6$,
J.Hamblen$^8$,
I.Harnarine$^6$,
M.Hauer$^2$,
C.Henderson$^4$,
D.J.Hofman$^6$,
R.S.Hollis$^6$,
R.Ho\l y\'{n}ski$^3$,
B.Holzman$^2$,
A.Iordanova$^6$,
E.Johnson$^8$,
J.L.Kane$^4$,
N.Khan$^8$,
P.Kulinich$^4$,
C.M.Kuo$^5$,
W.Li$^4$,
W.T.Lin$^5$,
C.Loizides$^4$,
S.Manly$^8$,
A.C.Mignerey$^7$,
R.Nouicer$^2$,
A.Olszewski$^3$,
R.Pak$^2$,
C.Reed$^4$,
E.Richardson$^7$,
C.Roland$^4$,
G.Roland$^4$,
J.Sagerer$^6$,
H.Seals$^2$,
I.Sedykh$^2$,
C.E.Smith$^6$,
M.A.Stankiewicz$^2$,
P.Steinberg$^2$,
G.S.F.Stephans$^4$,
A.Sukhanov$^2$,
A.Szostak$^2$,
M.B.Tonjes$^7$,
A.Trzupek$^3$,
C.Vale$^4$,
G.J.van~Nieuwenhuizen$^4$,
S.S.Vaurynovich$^4$,
R.Verdier$^4$,
G.I.Veres$^4$,
P.Walters$^8$,
E.Wenger$^4$,
D.Willhelm$^7$,
F.L.H.Wolfs$^8$,
B.Wosiek$^3$,
K.Wo\'{z}niak$^3$,
S.Wyngaardt$^2$,
B.Wys\l ouch$^4$\\
\vspace{3mm}
\footnotesize
%
%
%
%
$^1$~Argonne National Laboratory, Argonne, IL 60439-4843, USA\\
$^2$~Brookhaven National Laboratory, Upton, NY 11973-5000, USA\\
$^3$~Institute of Nuclear Physics PAN, Krak\'{o}w, Poland\\
$^4$~Massachusetts Institute of Technology, Cambridge, MA 02139-4307, USA\\
$^5$~National Central University, Chung-Li, Taiwan\\
$^6$~University of Illinois at Chicago, Chicago, IL 60607-7059, USA\\
$^7$~University of Maryland, College Park, MD 20742, USA\\
$^8$~University of Rochester, Rochester, NY 14627, USA\\
}


\begin{abstract}
The first measurement of pseudorapidity ($\Delta\eta$) and azimuthal angle ($\Delta\phi$) correlations between high transverse momentum charged hadrons ($p_T > 2.5$~GeV/c) and all associated particles is presented at both short- (small $\Delta\eta$) and long-range (large $\Delta\eta$) over a continuous pseudorapidity acceptance ($-4<\Delta\eta<2$).  In these proceedings, the various near- and away-side features of the correlation structure are discussed as a function of centrality in Au+Au collisions measured by PHOBOS at $\sqrt{s_{_{NN}}}=200$~GeV.  In particular, this measurement allows a much more complete determination of the longitudinal extent of the ridge structure, first observed by the STAR collaboration over a limited $\eta$ range.  In central collisions the ridge persists to at least $\Delta\eta=4$, diminishing in magnitude as collisions become more peripheral until it disappears around $N_{part}=80$.
\end{abstract}

\section*{Introduction}

One of the most fundamental discoveries at RHIC is that partons seem to interact strongly as they traverse the produced medium.  This has been seen in the single-particle spectra where the yields in central collisions at high $p_T$ are suppressed by a factor of five compared to binary scaling of p+p collisions \cite{PHOBOS_RAA,PHENIX_RAA,STAR_RAA}.  It is also seen in azimuthal correlations where back-to-back high $p_T$ particles disappear in central Au+Au collisions \cite{STAR_BackToBackJets}.  In fact, for high momentum associated particles the behavior is consistent with the surface emission of jets due the presence of an opaque medium that completely absorbs jets directed at the interior.

However, the energy and momentum of the away-side jet must be present in the final state, motivating the study of correlations between high $p_T$ triggers and lower $p_T$ associated particles.  At mid-rapidity, previous RHIC data show several novel features in the structure of such correlation functions: first, a broadening of the now re\"emergent away-side structure in $\Delta\phi$ relative to p+p \cite{PHENIX_LowPtAssoc,STAR_LowPtAssoc}, and second, the existence of an enhanced correlation near $\Delta\phi\approx0$ extending over several units of pseudorapidity -- this is what has been called the `ridge' \cite{STAR_Ridge}.  Although the ridge at mid-rapidity has been qualitatively described by a diverse assortment of proposed mechanisms \cite{Armesto,Hwa,Romantschke,Majumder,Shuryak,Pantuev,Wong}, the origin of the structure is still not well understood.  In this context, our goal is to use the uniquely broad acceptance of the PHOBOS detector to measure the ridge (and its dependence on event centrality) at large relative rapidity, and in so doing constrain the possible interpretations of particle production correlated with high $p_T$ trigger particles. 

\section*{Method}

The PHOBOS detector \cite{PHOBOS_NIM} consists of two spectrometer arms, which are used to select charged trigger tracks with $p_T>2.5$~GeV/c within the acceptance of $0<\eta^{trig}<1.5$.  Associated particles that escape the beam pipe ($p_T  > 4$~MeV/c at $\eta\approx3$, $p_T > 35$~MeV/c at $\eta\approx0$) are detected in a single layer of silicon (i.e. there is no $p_T$ selection) with the octagon subdetector ($|\eta|<3$).  Gaps in the octagon are filled using the first layers of the vertex and spectrometer detectors.  

\noindent The per-trigger conditional yield of charged particles,

\begin{equation}
\hspace{-0.4in}\frac{1}{N_{trig}} \frac{d^{2}N_{ch}}{d\Delta\phi d\Delta\eta}    =    B(\Delta\eta)  \cdot \left[\frac{s(\Delta\phi,\Delta\eta)}{b(\Delta\phi,\Delta\eta)}  - a(\Delta\eta) [ 1+2V(\Delta\eta) cos(2\Delta\phi) ]\right],
\label{eqn:CorrelatedYield}
\end{equation}

\noindent is calculated by taking the raw per-trigger distribution of same-event pairs, $s(\Delta\phi,\Delta\eta)$, and dividing by the raw mixed-event background distribution, $b(\Delta\phi,\Delta\eta)$, to remove random coincidences and acceptance effects.  This ratio is calculated in $1$~mm bins of vertex position along the beam, $v_z$, and averaged over the range $-15<v_z<10$~cm.  

Because elliptic flow is erased in the mixing of tracks and hits from different events, the remaining flow modulation, $V(\Delta\eta)$, which is approximated by the product of $\langle v_2^{trig}\rangle$ and $\langle v_2^{assoc}\rangle$, must also be subtracted.  The flow magnitude is calculated according to a parameterization based on published PHOBOS measurements of $v_2$ as a function of  $N_{part}$, $p_T$, and $\eta$, assuming a factorized form \cite{PHOBOS_Flow}.  The $v_2$ of trigger tracks is corrected for occupancy effects in the spectrometer, and the $v_2$ of associated hits is corrected for secondaries.  Both of these effects tend to suppress the magnitude of the observed $v_2$.  The scale factor $a(\Delta\eta)$, is introduced to account for the small difference in multiplicity between signal and mixed-event distributions.  It is calculated using the zero yield at minimum (ZYAM) method \cite{ZYAM} and is close to unity in all cases considered.  $B(\Delta\eta)$ is just the single-particle distribution ($dN/d\eta$) \cite{dNdeta} convoluted with the normalized $\eta$ distribution of trigger particles.  

The dominant systematic error in this analysis comes from the uncertainty in estimating the magnitude of $\langle v_2^{trig}\rangle\langle v_2^{assoc}\rangle$.  This flow uncertainty is typically on the order of 15-20\%, though in central collisions, where the subtracted flow is quite small compared to the resulting jet correlation, the uncertainty reaches 50\%.  

\section*{Results}

\begin{figure}[htbp]
   \centering
   \subfigure[p+p PYTHIA]{
   	\includegraphics*[width=.45\textwidth, viewport=20 10 550 360]{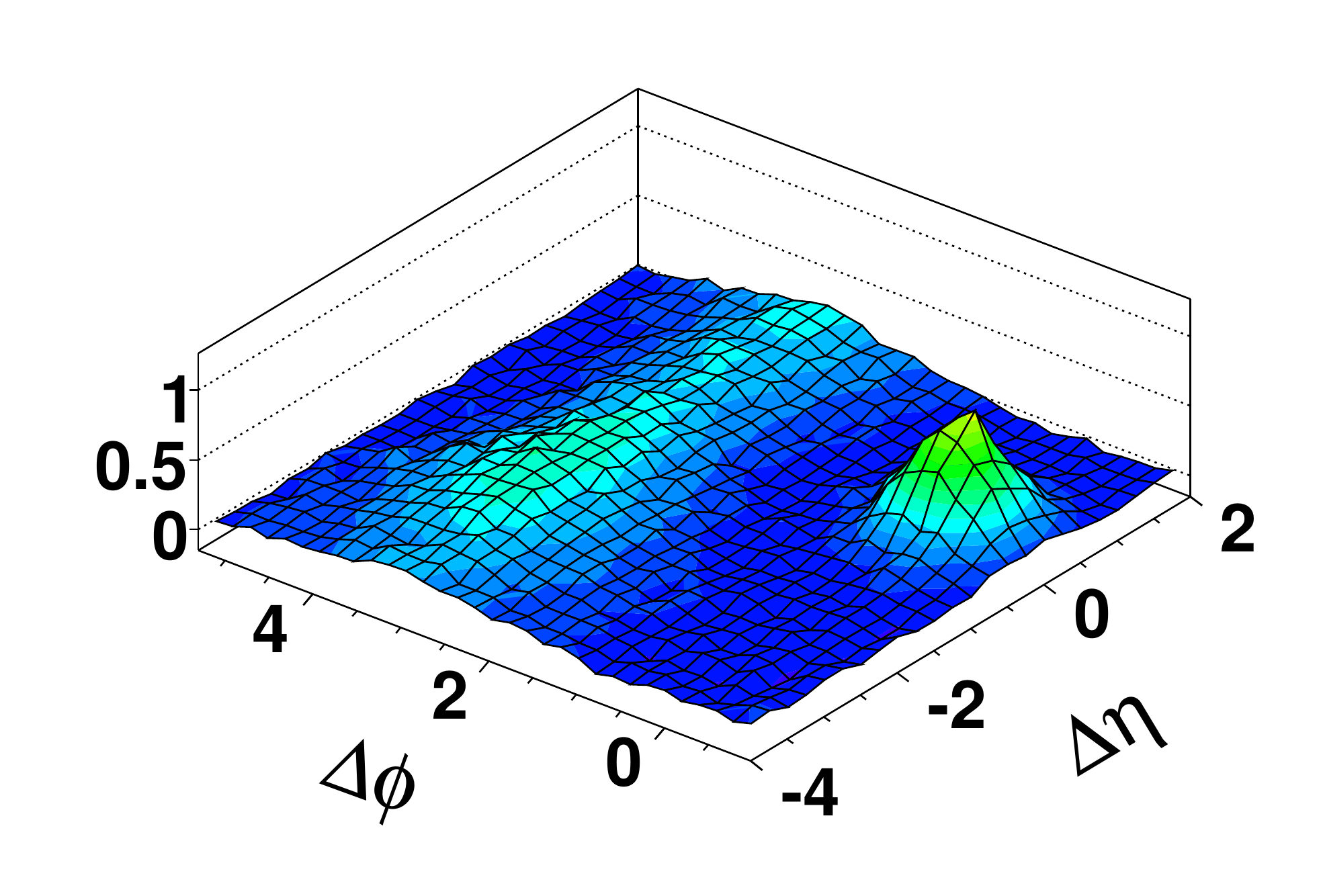} 
	\label{fig:pythiacorr}
   }
   \subfigure[Au+Au 0-30\%]{
   	\includegraphics*[width=.45\textwidth, viewport=20 10 550 360]{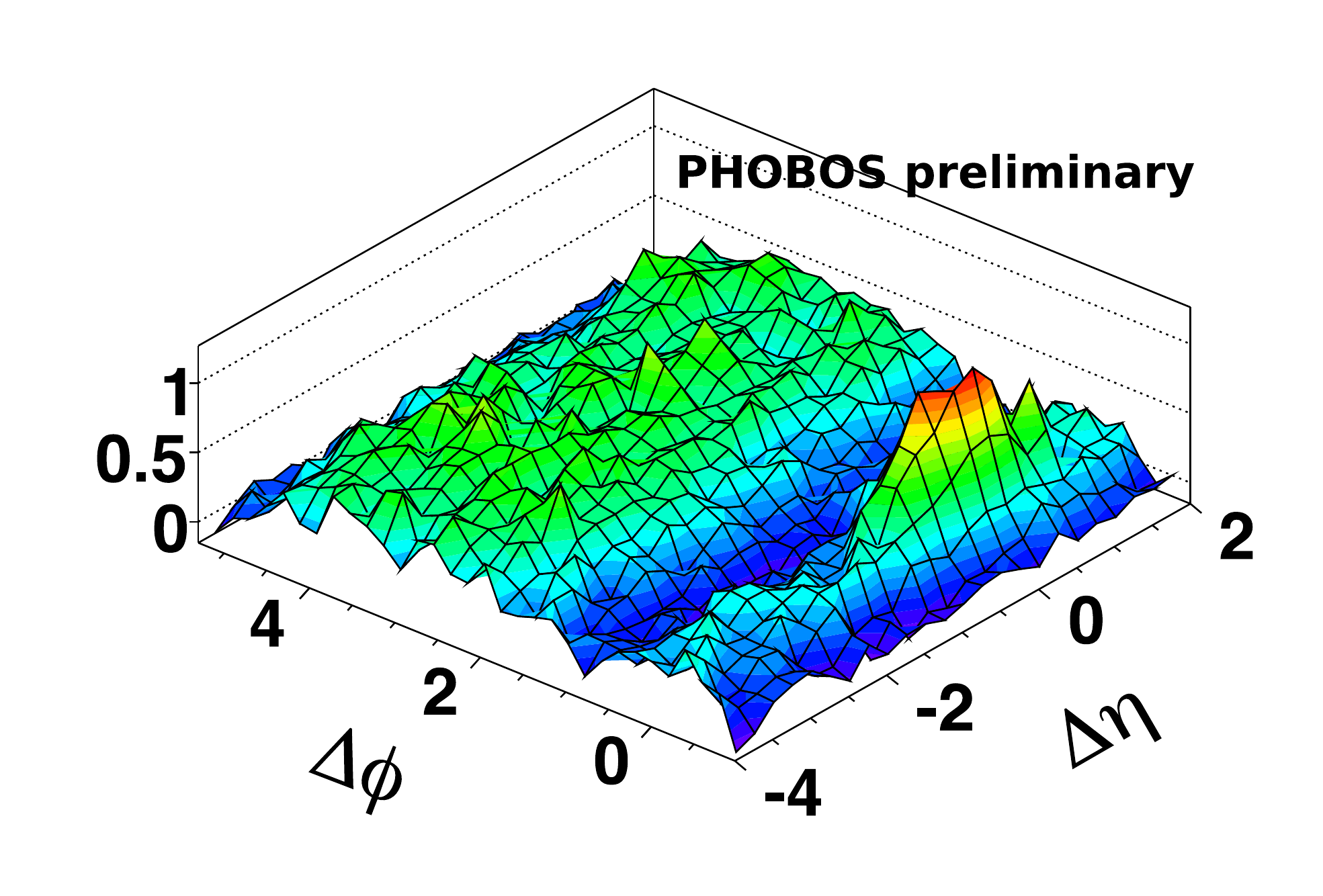} 
	\label{fig:auaucorr}
   }
   \caption{Per-trigger correlated yield with $p_T^{trig} >$~2.5 GeV/c as a function of $\Delta\eta$ and $\Delta\phi$ for 
   200 GeV \subref{fig:pythiacorr} PYTHIA p+p and \subref{fig:auaucorr} PHOBOS Au+Au collisions.}
   \label{fig:corrsurf}
\end{figure}

To understand the effects of the hot, dense medium on correlated particle production, the PHOBOS Au+Au data is compared to p+p events simulated with PYTHIA \cite{Pythia}.  The prominent features of the p+p correlation, shown in Fig.~\ref{fig:pythiacorr}, are a jet-fragmentation peak centered about $\Delta\eta\approx\Delta\phi\approx0$ and an away-side structure centered at $\Delta\phi \approx \pi$ that is similarly narrow in $\Delta\phi$ but extended in $\Delta\eta$ (since the hard scattering can involve partons with very different fractions of the proton momentum).  

In central Au+Au collisions, particle production correlated with a high $p_T$ trigger is strongly modified as shown in Fig.~\ref{fig:auaucorr}.  Not only is the away-side yield much broader in $\Delta\phi$, the near-side peak now sits atop an unmistakable ridge of correlated partners extending continuously and undiminished all the way to $\Delta\eta=4$.  

\begin{figure}[htbp]
   \centering
   \subfigure[]{
   	\includegraphics*[width=.5\textwidth, viewport=10 0 550 360]{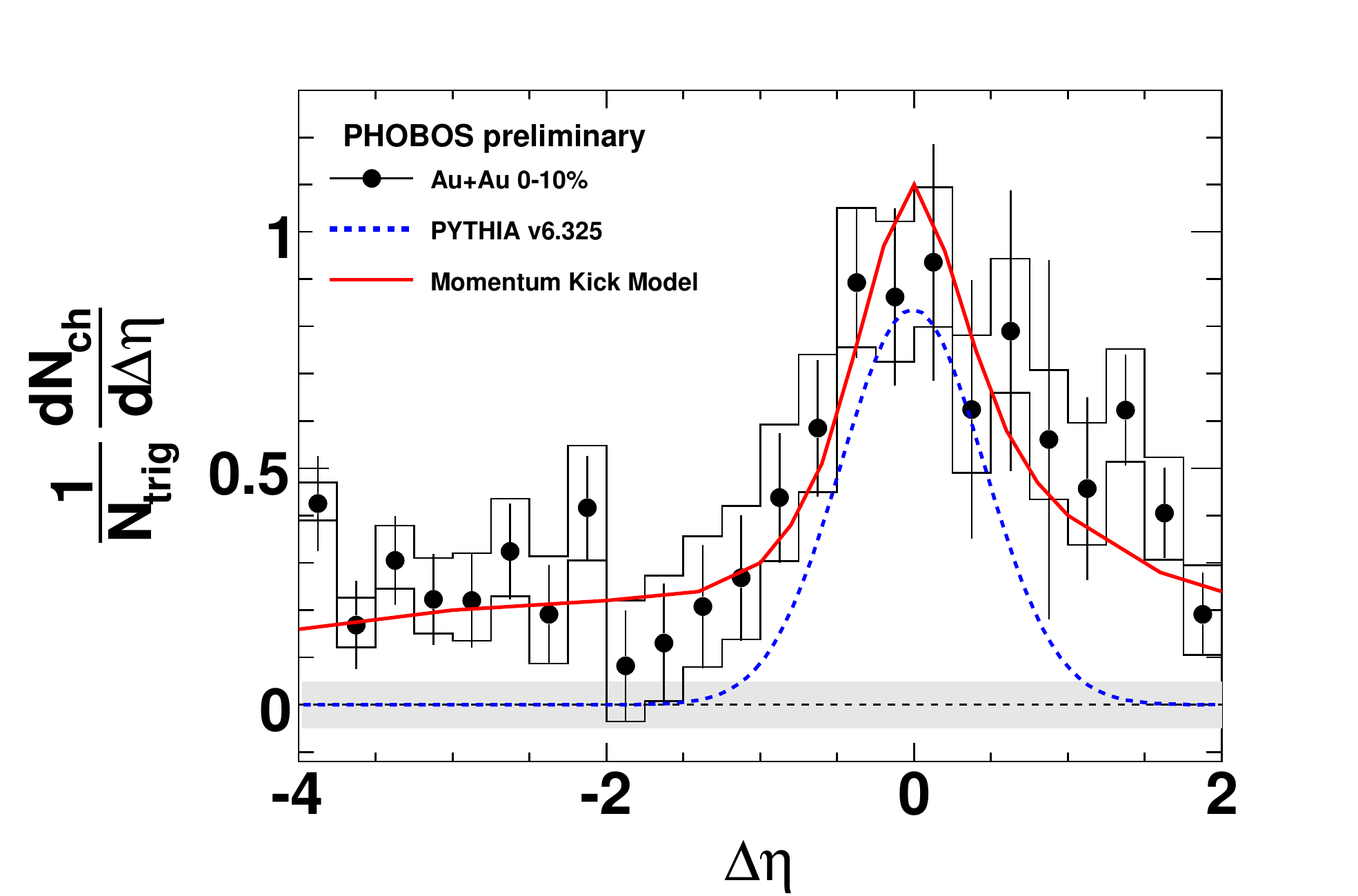} 
	\label{fig:NearSideVersusDeta}
	}
   \subfigure[]{
   	\includegraphics*[width=.42\textwidth, viewport=10 10 550 430]{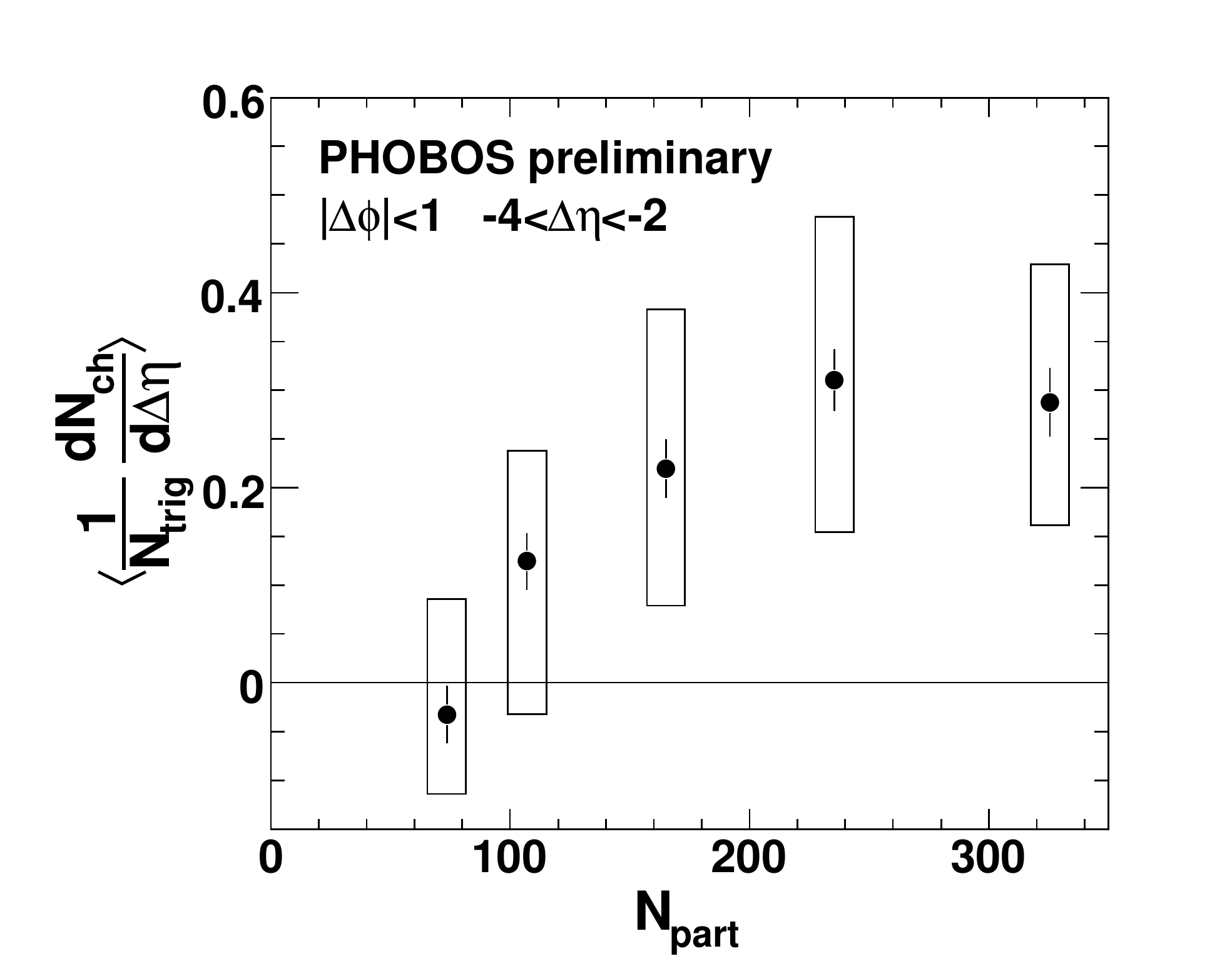} 
	\label{fig:RidgeVersusNpart}
	}
   \caption{\subref{fig:NearSideVersusDeta} Per-trigger correlated yield for 0-10\% central Au+Au integrated over the near-side ($|\Delta\phi| < 1$) compared to Pythia (dashed line) and momentum kick model prediction \cite{WongPrediction} (solid line) as a function of $\Delta\eta$.  Systematic uncertainty from $v_2$ estimate is represented by boxes.  \subref{fig:RidgeVersusNpart} Average ridge yield as a function of $N_{part}$ in the range $-4 < \Delta\eta < -2$.  Boxes correspond to errors on the $v_2$ estimate and ZYAM procedure.}
\end{figure}

To examine the near-side structure more closely, the correlated yield is integrated over the region $|\Delta\phi|<1$ and plotted as a function of $\Delta\eta$ in Fig.~\ref{fig:NearSideVersusDeta}.  For the 10\% most central Au+Au collisions, there is a significant and relatively flat correlated yield of about 0.3 particles per unit pseudorapidity far from the trigger.  The prediction of the momentum kick model \cite{WongPrediction}, which has been tuned to STAR measurements with $|\eta^{assoc}|<1$ \cite{STAR_LowPtAssoc} and $2.7<|\eta^{assoc}|<3.9$ \cite{STAR_LongRange}, is found to agree very closely with our results.  This model assumes a parton rapidity distribution at the time of the jet-parton collision that is much broader than the rapidity distribution of final state particles.  This would suggest that the extent of the ridge is actually sensitive to the very earliest moments of the Au+Au collision \cite{WongRapidityWidths}.   

Finally, the centrality dependence of the average ridge yield far from the trigger ($-4 < \Delta\eta < -2$) is shown in Fig.~\ref{fig:RidgeVersusNpart}.  The ridge yield decreases as one goes towards more peripheral collisions; it is consistent with zero in the most peripheral bin analyzed (40-50\%).  While the systematic errors do not yet exclude a smooth disappearance of the ridge as one approaches p+p collisions, these preliminary data suggest the ridge may have already disappeared by $N_{part}=80$.

\section*{Summary}

In these proceedings, preliminary PHOBOS measurements of the ridge at small $\Delta\phi$ have been presented over a broad range of $\Delta\eta$.  The fact that in central collisions the ridge extends to at least four units of rapidity away from the trigger is a quantitative challenge to theories that strive to explain the nature of the jet-medium interaction.  More theoretical studies will be required to determine which proposed mechanisms are consistent with the broad extent of the ridge and its dependence on collision geometry.

\section*{Acknowledgements}
%
%
%
%
This work was partially supported by U.S. DOE grants 
DE-AC02-98CH10886,
DE-FG02-93ER40802, 
DE-FG02-94ER40818,  
DE-FG02-94ER40865, 
DE-FG02-99ER41099, and
DE-AC02-06CH11357, by U.S. 
NSF grants 9603486, 
0072204,            
and 0245011,        
by Polish MNiSW grant N N202 282234 (2008-2010),
by NSC of Taiwan Contract NSC 89-2112-M-008-024, and
by Hungarian OTKA grant (F 049823).


\end{document}